\documentclass[twoside]{article}
\usepackage{fleqn,espcrc2}
 
\usepackage{epsfig}
\usepackage{graphicx}
\usepackage{amsmath}
\usepackage[figuresright]{rotating}                                             

\def\lsi{\raise0.3ex\hbox{$<$\kern-0.75em\raise-1.1ex\hbox{$\sim$}}}
\def\gsi{\raise0.3ex\hbox{$>$\kern-0.75em\raise-1.1ex\hbox{$\sim$}}}

\title{Lattice QCD at finite T and $\mu$ and the critical point of QCD} 

\author{Z.~Fodor and S.D.~Katz
\address{Institute for Theoretical Physics, E\"otv\"os University, P\'azm\'any
1, H-1117 Budapest, Hungary}}

\begin{document}
\begin{abstract} 
We propose a method to study lattice QCD at finite T and $\mu$.
We compare it with direct results and with the Glasgow method 
by using $n_f$=4 QCD at Im($\mu$)$\neq$0. 
We locate the critical endpoint
(E) of QCD on the Re($\mu$)-T plane. In this study we use $n_f$=2+1 dynamical 
staggered quarks with semi-realistic masses on $L_t=4$ lattices.
\end{abstract}
\maketitle

{\it I. Introduction.---}
QCD at finite $T$ and $\mu$ is of fundamental importance,
since it describes relevant features of particle physics
in the early universe, in neutron stars and in heavy ion collisions.
Extensive experimental work has been done
with heavy ion collisions at CERN and Brookhaven to explore
the $\mu$-$T$ phase diagram. It is
a long-standing non-perturbative question, whether a critical point
exists on the $\mu$-$T$ plane,
and particularly how to tell its location theoretically
\cite{crit_point}.           
   
QCD at finite $\mu$ can be 
formulated on the lattice \cite{HK83}; however, standard 
Monte-Carlo techniques can not be used. The reason 
is that for Re($\mu$)$\neq$0 the determinant of 
the Euclidean Dirac operator is complex. This fact
spoils any importance sampling method. 

An attractive approach to alleviate the problem 
is the ``Glasgow method'' (see e.g. Ref. \cite{glasgow}) in which the 
partition function ($Z$) is expanded in powers of $\exp(\mu/T)$
by using an ensemble of configurations weighted by the $\mu$=0 action. 
After collecting more than 20 million configurations only unphysical
results were obtained. 
The reason is that the $\mu$=0 ensemble does not overlap enough 
with the finite density states of interest. 
A recent review on QCD at finite T-$\mu$ is Ref. \cite{H01}.

We propose a method 
to reduce the overlap problem and determine the
phase diagram in the $\mu$-T plane (for details see \cite{FK01}).  
We also locate the critical point of QCD. 
(Similar technique was successful for determining  
the endpoint of the hot electroweak plasma \cite{ewpt}
e.g. on 4D lattices.) 

{\it II. Overlap improving multi-parameter reweighting.---}
Let us study a generic system of fermions $\psi$ and bosons $\phi$,
where the fermion Lagrange density is ${\bar \psi}M(\phi)\psi$.
Integrating over the Grassmann fields we get:
\begin{equation*}
Z(\alpha)=\int{\cal D}\phi \exp[-S_{bos}(\alpha,\phi)]\det M(\phi,\alpha),
\end{equation*}
where $\alpha$ denotes a set of parameters of
the Lagrangian. In the case staggered QCD $\alpha$
consists of $\beta$,
the quark mass ($m_q$) and $\mu$.
For some choice of the
parameters $\alpha$=$\alpha_0$
importance sampling can be done (e.g. for Re($\mu$)=0).
Rewriting the above equation one obtains       
\begin{eqnarray*}\label{reweight}
Z(\alpha)=
\int {\cal D}\phi \exp[-S_{bos}(\alpha_0,\phi)]\det M(\phi,\alpha_0)&& 
\\
\left\{\exp[-S_{bos}(\alpha,\phi)+S_{bos}(\alpha_0,\phi)]
{\det M(\phi,\alpha)  \over \det M(\phi,\alpha_0)}\right\}.&&
\end{eqnarray*}
We treat the curly bracket as an observable
--which is measured on each configuration--
and the rest as the measure. It is known that changing
only one parameter of the ensemble
generated at $\alpha_0$ provides an accurate value for some observables
only for high statistics. This is ensured by 
rare fluctuations as the mismatched measure occasionally sampled the
regions where the integrand is large. This is the so-called
overlap problem. Having several parameters
the set $\alpha_0$ can be adjusted to ensure
a better overlap than obtained by varying only one parameter. 
                           
The basic idea of the method as applied to dynamical QCD can be 
summarized as follows. We study the system at ${\rm Re}(\mu)$=0 around 
its transition point. Using a Glasgow-type technique we calculate the 
determinants for each configuration for a set of $\mu$, which, similarly 
to the Ferrenberg-Swendsen method \cite{FS89}, can be used for 
reweighting.  The average plaquette values can be used to perform an 
additional reweighting in $\beta$.  Since transition configurations were 
reweighted to transition configurations a much better overlap can be 
observed than by reweighting pure hadronic configurations to transition 
ones as done by the Glasgow-type techniques (moving along the transition 
line was also suggested by Ref. \cite{AKW99}).

{\it III. Illustration and direct test.---}
We have directly tested these ideas in four-flavor QCD 
with $m_q$=0.05 dynamical 
staggered quarks. 

\begin{figure}
\epsfig{file=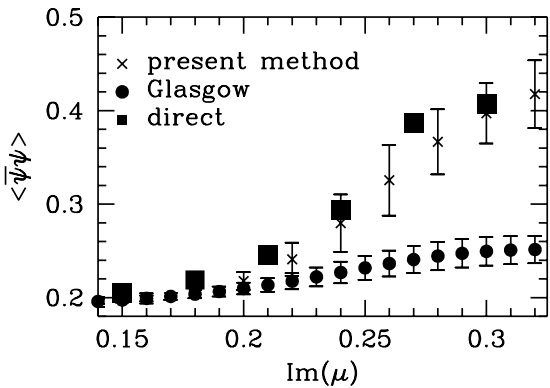,bb= 17 285 570 610, width=7.4cm}\label{im_mu}
\caption{The average of ${\bar \psi}\psi$ at $\beta$=5.085 as a 
function of Im($\mu$), for direct results (squares; their sizes give 
the errors), our technique (crosses) and Glasgow-type reweighting (dots). 
}
\end{figure}

We first collected 1200 independent V=4$\cdot 6^3$ configurations at 
Re($\mu$)=Im($\mu$)=0 and some $\beta$ 
values and used the Glasgow-reweighting and 
also our technique to study Re($\mu$)=0, Im($\mu$)$\neq$0. At 
Re($\mu$)=0, Im($\mu$)$\neq$0 direct simulations are possible. 
After performing these direct simulations as well, a clear 
comparison can be done. Figure 1 shows the predictions of 
the three methods for the average quark condensates at $\beta$=5.085
as a function of Im($\mu$).
The predictions of our method agree with the direct results,
whereas for larger Im($\mu$) the predictions of the Glasgow
method are by several standard deviations off. 
Based on these experiences we expect that our
method can be successfully applied at Re($\mu$)$\neq$0. 

{\it IV. $n_f\neq$4 staggered quarks. }---  
In QCD with $n_f$ staggered quarks
one should change the determinants to their $n_f$/4 power in our two 
equations. Importance sampling works also in this case  at some $\beta$ and 
at Re($\mu$)=0. Since $\det M$ is complex, one should choose 
among the Riemann-sheets of the fractional power. We solved this problem
analytically \cite{FK01}.

\begin{figure}
\epsfig{file=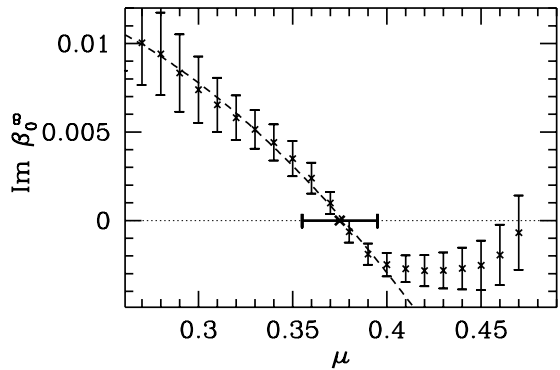,bb= 17 297 570 610,width=7.4cm} \label{infV}
\caption{Im($\beta_0^\infty$) as a function of the chemical potential.}
\end{figure}

In the 
following we keep $\mu$ real and look for the zeros of $Z$ on the 
complex $\beta$ plane.  At a first order phase transition the free 
energy $\propto \log Z(\beta)$ is non-analytic. 
Clearly, a 
phase transition appears only in the V$\rightarrow \infty$ limit, 
but not in a finite $V$. Nevertheless, $Z$
has zeros at finite V, generating the non-analyticity of the 
free energy, the Lee-Yang zeros \cite{LY52}. 
These are at complex values
of the parameters, in our case at complex $\beta$. For a 
system with a first order transition these zeros
approach the real axis in the V$\rightarrow \infty$ limit
(detailed analysis suggests $1/V$ scaling).   
This V$\rightarrow \infty$ limit generates the non-analyticity of
the free energy. For a system with crossover  
$Z$ is analytic, and the zeros do
not approach the real axis in the V$\rightarrow \infty$ limit.

At T$\neq$0 we used $L_t$=4, $L_s$=4,6,8 lattices. T=0 runs were done on
$10^3\cdot$ 16 lattices. $m_{u,d}$=0.025 and $m_s$=0.2 were
our bare quark masses. 

At  $T\neq 0$ we determined the complex valued Lee-Yang zeros, 
$\beta_0$, for different V-s as a function of $\mu$. Their 
V$\rightarrow \infty$ limit was given by a $\beta_0(V)=\beta_0^\infty+\zeta/V$
extrapolation. We used 14000, 3600 and 840 configurations on 
$L_s$=4,6 and $8$ lattices, respectively. Figure 2 
shows Im($\beta_0^\infty$) as a function of $\mu$.  For small
$\mu$-s the extrapolated Im($\beta_0^\infty$) is inconsistent with
a vanishing value, and predicts a crossover.
Increasing $\mu$ the value of Im($\beta_0^\infty$) decreases, 
thus the transition becomes consistent with a first order phase
transition. (Note, that systematic overshooting is a finite V effect.)
The statistical error was determined by jackknife samples of the
total $L_s=4,6$ and $8$ ensembles. 
Our primary result is $\mu_{end}=0.375(20)$. 

To set the physical scale we used a 
weighted average of $R_0$,  $m_\rho$  and 
$\sqrt{\sigma}$. 
Note, that (including systematics due to 
finite V) we have 
$(R_0\cdot m_\pi)=0.73(6)$, which is at least twice, $m_{u,d}$ is
at least four times
as large as the physical values. 

Figure 3 shows the phase diagram in
physical units, thus
$T$ as a function of $\mu_B$, the baryonic chemical potential 
(which is three times larger then the quark chemical potential). 
The endpoint
is at $T_E=160 \pm 3.5$~MeV, $\mu_E=725 \pm 35$~MeV.
At $\mu_B$=0 we obtained $T_c=172 \pm 3$~MeV. 

\begin{figure}
\epsfig{file=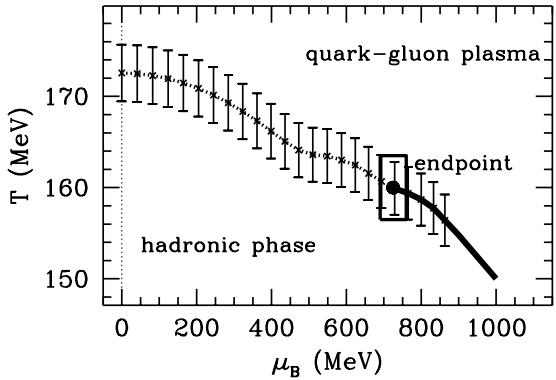,bb= 17 297 570 610,width=7.4cm}\label{physical}
\caption{
The T-$\mu$ diagram. Direct results are given with errorbars.
Dotted line shows the crossover, solid line the first order 
transition. The  box gives the uncertainties of the endpoint.}
\end{figure}

{\it IV. Conclusions.}---
We proposed a method --an overlap improving multi-parameter reweighting 
technique-- to numerically study non-zero $\mu$ and determine the 
phase diagram in the $T$-$\mu$ plane.   
Our method is applicable to any number of Wilson or staggered quarks. 
As a direct test we showed that for Im($\mu$)$\neq$0 the predictions 
of our method are 
in complete agreement with the direct simulations, whereas the Glasgow
method suffers from the well-known overlap problem.

We studied the $\mu$-$T$ phase diagram of QCD with 
dynamical $n_f$=2+1 quarks. 
Using our method we obtained 
$T_E$=160$\pm$3.5~MeV and $\mu_E$=725$\pm$35~MeV for the endpoint. 
Though $\mu_E$ is too
large to be studied at RHIC/LHC, the endpoint would 
probably move closer to the $\mu$=0 axis 
when the quark masses get reduced. 
At $\mu$=0 we obtained $T_c$=172$\pm$3~MeV.
Clearly, more work is needed to get
the final values. One has to  extrapolate to zero step-size
in the R-algorithm and to the thermodynamic, chiral and continuum limits.    

This work was partially supported by Hung. Sci. 
grants No. 
OTKA-\-T34980/\-T29803/\-T22929/\-M28413/\-OM-MU-708/\-IKTA111/\-NIIF. 
This work was in part based 
on the MILC collaboration's public lattice gauge theory code:
http://physics.indiana.edu/\~{ }sg/milc.html.

\end{document}